\documentclass[prd,twocolumn,showpacs,amsmath,amssymb,nofootinbib,superscriptaddress]{revtex4}
\usepackage{graphicx}
\usepackage{epsfig}
\usepackage{bm}
\usepackage{amsfonts}
\usepackage[latin1]{inputenc}
\usepackage{amssymb}
\usepackage{color}
\usepackage{float}
\usepackage{amsmath}                  
\usepackage{dcolumn}                
\usepackage{hyperref}
\usepackage{array}
\usepackage{lscape}

\def\be{\begin{equation}}
\def\ee{\end{equation}}
\def\bea{\begin{eqnarray}}
\def\eea{\end{eqnarray}}
\def\eqi{\begin{equation}}
\def\eqf{\end{equation}}
\def\eqia{\begin{eqnarray}}
\def\eqfa{\end{eqnarray}}
\def\lcdm{$\Lambda$CDM }

\begin{document}

\title{Constraining the dark energy models with H(z) data: an approach independent of $H_{0}$}

\author{Fotios K. Anagnostopoulos}\affiliation{National and Kapodistrian University of Athens, Physics Department, Panepistimioupoli Zografou,  15772, Athens, Greece}\email{fotis-anagnostopoulos@hotmail.com}
\author{Spyros Basilakos}
\affiliation{Academy of Athens, Research Center for Astronomy and
Applied Mathematics, Soranou Efesiou 4, 11527, Athens, Greece}\email{svasil@academyofathens.gr}


\begin{abstract}
We study the performance of the latest 
$H(z)$ data 
in constraining the cosmological parameters of different cosmological models, 
including that of Chevalier-Polarski-Linder $w_{0}w_{1}$ parametrization. 
First, we introduce a statistical procedure in which the chi-square estimator 
is not affected by the value of the Hubble constant. 
As a result, we find that the $H(z)$ data do not rule out the possibility 
of either non-flat models or dynamical dark energy cosmological models. 
However, we verify that the time varying equation-of-state parameter $w(z)$ 
is not constrained by the current expansion data. 
Combining the $H(z)$ and the Type Ia supernova data we find 
that the $H(z)$/SNIa overall statistical analysis provides a 
substantial 
improvement of the cosmological constraints with respect to those of the 
$H(z)$ analysis. Moreover, the $w_{0}-w_{1}$ parameter space provided by 
the $H(z)$/SNIa joint analysis is in a very good agreement 
with that of Planck 2015, which confirms that the present  
analysis with the $H(z)$ and SNIa probes correctly reveals 
the expansion of the Universe as found by the team of Planck.
Finally, we generate sets of Monte Carlo realizations in order 
to quantify the ability of the $H(z)$ data to provide strong constraints 
on the dark energy model parameters. 
The Monte Carlo approach shows significant improvement of 
the constraints, when increasing the sample to 100 $H(z)$ measurements. 
Such a goal 
can be achieved in the future, especially in the 
light of the next generation of surveys.

\end{abstract}

\maketitle


\section{Introduction}

The general picture of the Cosmos, as it is established by the analysis of the recent cosmological data  
(see \cite{Planck16} and references therein), is described with a cosmological scenario that consists 
$\sim 30\%$ of matter (baryonic and dark) and the rest corresponds to the so called 
dark energy (DE). This mysterious component of the cosmic fluid 
plays an eminent role in cosmological studies because it is responsible 
for the accelerated expansion of the Universe.
Also, current observations seem to favor an 
isotropic, homogeneous and spatially flat 
universe.

During  the  last  decades, different classes of theoretical models have been 
introduced in order to explain the accelerating Universe\footnote{for a review, see \cite{Amendola}.}, 
giving rise to a scholastic debate about what is the exact description and the key points of each scheme. One of the fundamental questions of modern cosmology that subsequently emerges is what is the model that best describes the accelerated expansion of 
the universe, \cite{maartens2010}. 
A prominent path in order to distinguish 
the various cosmological models 
is to probe the cosmic history \cite{Probes} of the universe, using 
either the luminosity distance of standard candles
or the angular diameter distance of standard rulers.


In general, the
geometrical probes used to map the cosmic expansion history
involve a combination of standard candles 
(SNIa \cite{Suzuki:2011hu,Betoule14}), GRBs \cite{GRB}, HII \cite{Plionis2011,Chavez2016}), standard rulers
(clusters, CMB sound horizon detected through Baryon Acoustic
Oscillations (BAO); \cite{Blake,Alam2016}), the CMB 
angular power spectrum \cite{Planck16}
and  recently,  data from gravitational wave measurements, 
the so called 'standard sirens', \cite{Calabrese2016}. 
Alternatively,
dynamical probes of the expansion history based on measures of the
growth rate of matter perturbations (for recent studies see \cite{Nes17} 
and references therein) are also used towards tracing the cosmic 
expansion and they are confined to
relatively low redshifts similar to those of Type Ia supernova data 
$z\simeq 1.4$. 
The aforementioned observations probe
the integral of the Hubble parameter $H(z)$, hence they 
give us indirect information for the cosmic expansion.
Also, it is worth noting that in some cases the data suffer 
from the so called circularity
problem, the fact that one needs to impose a fiducial cosmology in order 
to be able to define the data (see for example 
\cite{Wang2005}, \cite{Shapiro2006}).

Among the large body of cosmological data the only data-set 
that provides a direct measurement of the cosmic expansion 
is the $H(z)$ sample and indeed a 
plethora of papers have been published 
(e. g. \cite{Samushia2006}, \cite{Farooq13}, \cite{ChimentoRicharte2013}, \cite{Ferreiraetal2013}, \cite{Capozzielloetal2014}, \cite{GruberLuongo14}, \cite{Forte2014}, \cite{Dankiewiczetal2014}, \cite{Caietal2015},  \cite{Melia2015}, \cite{Chenetal2016a}, \cite{MukherjeeBanerjee2016}, \cite{Farooq16}, \cite{Saridakis}, \cite{Cardasian}, \cite{Saridakisf(R)}) 
which determine the dynamical 
characteristics of various DE cosmological models, including those 
of modified gravity.  
Today, the most recent $H(z)$ data 
trace the cosmic expansion rate up to
redshifts of order $z\simeq 2.4$, while 
there are proposed methods \cite{SantageLoeb} 
which potentially could expand the $H(z)$ measurements to $z\le 5$.
As expected using the $H(z)$ data in constraining the cosmological models 
via the standard likelihood analysis, 
one has to deal with the 
Hubble constant, namely $H_{0}$. However, the best choice of the value of 
$H_{0}$ is rather uncertain. Indeed, several studies 
on the determination of the
Hubble constant have indicated a $\sim 3.1 \sigma$ tension 
between the value obtained by the Planck team (see \cite{Planck16}), namely   
$H_{0} = 67.8 \pm 0.9$ Km/s/Mpc and the results provided by the  
SNIa project (Riess \emph{et al}. \cite{Riess16}) of 
$H_{0} = 73.24 \pm 1.74$ Km/s/Mpc.
In order to alleviate this problem we propose in the current work 
a statistical method which is not affected 
by the value of $H_{0}$.  


The structure of the article is as follows: 
In Sec. \ref{Methodology} we present the $H(z)$ data used and the 
related statistical analysis.
At the beginning of Sec. \ref{Cosmology} we describe the main properties 
of the most basic 
DE models  
and then we focus on the cosmological constrains.
In Sec. \ref{Strategy} we discuss the Monte Carlo simulations 
used towards planning future $H(z)$ measurements in order to 
place better constraints on the DE model parameters. 
Finally, in Sec. \ref{Discussion} we provide a detailed 
discussion of our results and 
we summarize our conclusions in Sec.  \ref{Conclusions}.

\begin{table}[h!]
\caption{
The observational data-set that was used in this paper. The data-set, compiled by Farooq \emph{et al}, 2016 \cite{Farooq16} consists of $N = 38$ observations.
}
 \label{tab:farooq_data}
 \centering
\begin{tabular}{ ||c|c|c|c|| }

 \hline
 $z$ & $H(z) [{\rm Km /s/Mpc}]$ & $\sigma_{H}[{\rm Km /s/Mpc}]$ & Method/Ref.\\
\hline 
 0.070 &  69.0 & 19.6 & \cite{Zhang2014}\\
0.090 & 69.0 & 12.0 &  \cite{Simon2005}\\
0.120 & 68.6 & 26.2 &  \cite{Zhang2014}\\
0.170 & 83.0 & 8.0 &  \cite{Simon2005}\\
0.179 & 75.0 & 4.0 & \cite{Moresco2012}\\
0.199 & 75.0 & 5.0 & \cite{Moresco2012}\\
0.200 & 72.9 & 29.6 & \cite{Zhang2014}\\
0.270 & 77.0 & 14.0 &\cite{Simon2005}\\
0.280 & 88.8 & 36.6 &\cite{Zhang2014}\\ 
0.352 & 83.0 & 14.0 &\cite{Moresco2012}\\
0.380 & 81.5 & 1.9 &  \cite{Alam2016} \\ 
0.3802 & 83.0 & 13.5 &  \cite{Moresco2016}\\
0.400 & 95.0 & 17.0 &  \cite{Simon2005}\\
0.4004 & 77.0 & 10.2 &  \cite{Moresco2016}\\
0.4247 & 87.1 & 11.2 &   \cite{Moresco2016}\\
0.440 & 82.6 & 7.8 &  \cite{Blake2012}\\
0.4497 & 92.8 & 12.9 &   \cite{Moresco2016}\\
0.4783 & 80.9 & 9.0 &  \cite{Moresco2016}\\
0.480 & 97.0 & 62.0 &  \cite{Stern2010}\\
0.510 & 90.4 & 1.9 &  \cite{Alam2016} \\
0.593 & 104.0 & 13.0 & \cite{Moresco2012}\\
0.600 & 87.9 & 6.1 & \cite{Blake2012}\\
0.610 & 97.3 & 2.1 &  \cite{Alam2016} \\
0.680 & 92.0 & 8.0 & \cite{Moresco2012}\\
0.730 & 97.3 & 70.0 &  \cite{Blake2012}\\
0.781 & 105.0 & 12.0 & \cite{Moresco2012}\\
0.875 & 125.0 & 17.0 & \cite{Moresco2012}\\
0.880 & 90.0 & 40.0 &  \cite{Stern2010}\\
0.900 & 117.0 & 23.0 & \cite{Simon2005}\\
1.037 & 154.0 & 20.0 & \cite{Moresco2012}\\
1.300 & 168.0 & 17.0 & \cite{Simon2005}\\
1.363 & 160.0 & 33.6 &  \cite{Moresco2015}\\
1.430 & 177.0 & 18.0 & \cite{Simon2005}\\
1.530 & 140.0 & 14.0 & \cite{Simon2005}\\
1.750 & 202.0 & 40.0 & \cite{Simon2005}\\
1.965 & 186.5 & 50.4 & \cite{Moresco2015}\\
2.340 & 222.0 & 7.0 &  \cite{Delubac2015}\\
2.360 & 226.0 & 8.0 & \cite{Font-Ribera2014}\\   

 \hline

\end{tabular}
\end{table}


\begin{figure}[ht]
\label{results1}
\includegraphics[width=0.5\textwidth]{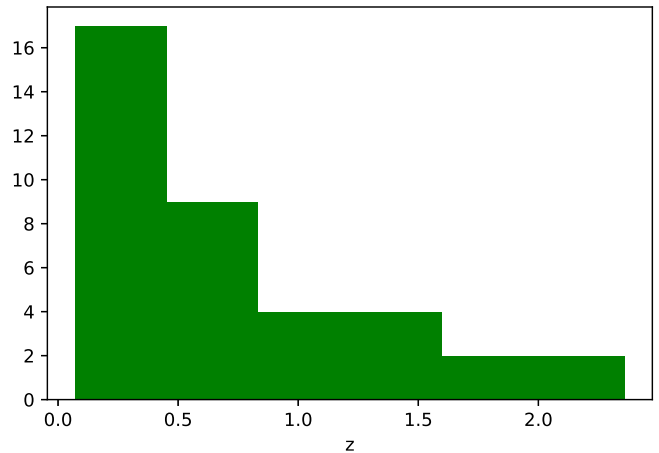}
\includegraphics[width=0.5\textwidth]{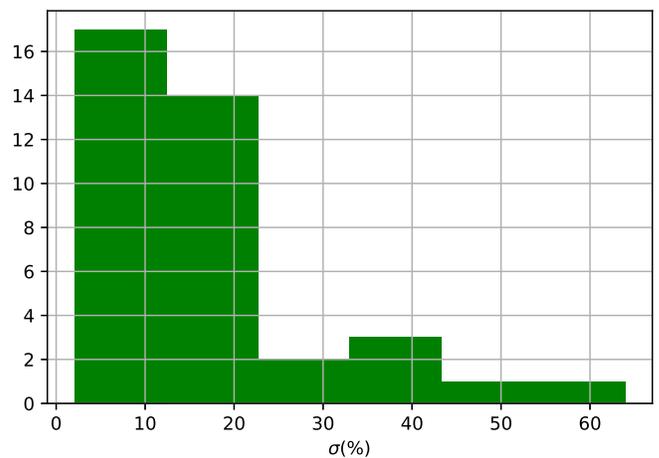}
\caption{The redshift (upper panel) and the 
relative error (lower panel) distributions of our dataset.}
\end{figure}

\section{Statistical analysis with H(z) data}
\label{Methodology}

In this section we discuss the details of the statistical
analysis and on the observational sample that we utilize
in order to place constraints on the cosmological parameters.
In particular, we use the cosmic expansion data as collected 
by Farooq \emph{et al.} \cite{Farooq16} (see Table I and the corresponding references) 
for which the Hubble parameter is available as a function of redshift.
Notice, that the $H(z)$ sample contains 38 entries in the following redshift range 
$0.07\leq z\leq 2.36$.  In Fig.1, we present 
the normalized redshift distribution of the 
$H(z)$ data and the corresponding distribution of the 
relative uncertainty $\sigma(\%)=\sigma_{H(z_{i})}/H(z_{i})$.
Also, we find no significant correlation between
$\sigma$ and redshift in that range. 

First let us assume that we have a dark energy model that
includes $n$-free parameters, provided by
the statistical vector $\phi^{\mu}=(\phi^{0},\phi^{1},...\phi^{n})$.
In order to put constraints on $\phi^{\mu}$ we need to 
implement a standard $\chi^2$-minimization procedure,
which in our case is written as 
\begin{eqnarray}
\label{eq:xtetr}
\chi^2(\phi^\mu)=\sum_{i=1}^{N} \left[\frac{H_{D}(z_{i})-H_{M}(z_{i},\phi^{\mu})}{\sigma_{i}}\right]^2
\end{eqnarray}
where $H_{D}(z_{i})$, $\sigma_{i}$ are the observational data and 
the corresponding uncertainties at the observed redshift, $z_{i}$. 
The capital letters $M$ and $D$ stand 
for model and data respectively. 
In this case 
the theoretical Hubble parameter is written as  
\begin{eqnarray}
\label{eq1}
H_{M}(z,\phi^{\mu}) = H_{0} E(z,\phi^{\mu+1})
\end{eqnarray}
where $H_{0}$ is the current value of Hubble parameter, namely the 
Hubble constant, $E(z)$ is the normalized Hubble function 
and the vector $\phi^{\mu}$ contains the cosmological parameters. 
In this framework we observe that the statistical vector becomes 
$\phi^{\mu}=(H_{0},\phi^{\mu+1})$, where 
the components $\phi^{\mu+1}$ contains the
free parameters which are related with the matter density, spatial curvature 
and dark energy.

Therefore, in order to proceed with the statistical analysis 
we need to either know the exact value of the Hubble constant 
or having it as a free parameter.
The most recent results on the determination of the
Hubble constant have found a 
$\sim 3.1 \sigma$ tension 
between the value obtained by SN Ia project
(Riess et al. \cite{Riess16}) of $H_{0} = 73.24 \pm 1.74$ Km/s/Mpc
and the results from Planck  
(see \cite{Planck16}) 
of $H_{0} = 67.8 \pm 0.9$ Km/s/Mpc.
The Hubble constant problem has inspired us to 
propose a technique which provides the chi-square 
estimator independent from the value of $H_{0}$.  
At this point we present the basic ingredients 
towards marginalizing $\chi^{2}$ over $H_{0}$.\footnote{Similar analysis has been proposed 
by Taddei \& Amendola \cite{Tad} and Basilakos \& Nesseris \cite{BasilakosNesseris2016}) 
in order to marginalize chi-square function of the growth rate data over the value of the rms fluctuations 
at $8h^{-1}$Mpc, namely $\sigma_{8}$.} 
Indeed, inserting \eqref{eq1} into \eqref{eq:xtetr} the latter equation simply becomes 

\begin{eqnarray}
\chi^2(\phi^{\mu})=A H_{0}^2-2BH_{0}+\Gamma,
\end{eqnarray}
where
\begin{eqnarray*}
A=\sum_{i=1}^{N}\frac{E^2(z_{i})}{\sigma_{i}^2}
\end{eqnarray*}

\begin{eqnarray*}
B=\sum_{i=1}^{N}\frac{E(z_{i})H_{D}(z_{i})}{\sigma_{i}^2}
\end{eqnarray*}

\begin{eqnarray*}
\Gamma=\sum_{i=1}^{N} \frac{H_{D}(z_{i})^2}{\sigma_{i}^2}
\end{eqnarray*}

In this context the corresponding likelihood function is written as 

\begin{eqnarray}
\mathcal{L}=e^{-x^2/2}\Rightarrow \mathcal{L}={\rm exp}\left[\frac{A H_{0}^2-2BH_{0}+\Gamma}{2} \right]
\end{eqnarray}
or
\begin{eqnarray*}
\mathcal{L}(D|\phi^{\mu},M)=
{\rm exp}\left[\frac{A\left(H_{0}-\frac{B}{A}\right)^2-\frac{B^2}{A}+\Gamma}{2}\right]\;.
\end{eqnarray*}
Using Bayes's theorem and marginalizing over $H_{0}$ we arrive at 
\begin{eqnarray}
p(\phi^{\mu}|D,M)=\frac{1}{p(D|M)}\int e^{-\frac{A(H_{0}-B/A)^2-B^2/A+\Gamma}{2}}dH_{0} \;.
\end{eqnarray}
Furthermore, considering that $H_{0}$ lies in the range 
$H_{0} \in (0,+\infty)$, 
introducing the variable $y=H_{0}-B/A$ and utilizing 
flat priors $p(\phi^{\mu}|M,H_{0})= 1$ we obtain after some simple 
calculations 

\begin{eqnarray}
\label{action11}
p(\phi^{\mu}|D,M)=\frac{1}{p(D|M)}e^{-\frac{1}{2}\left(\Gamma-B^2/A\right)}\int_{-\frac{B}{A}}^{+\infty}e^{-\frac{A}{2}y^2}dy 
\end{eqnarray}
or
\begin{equation}
p(\phi^{\mu}|D,M)=
\frac{1}{p(D|M)}  e^{ -\frac{1}{2} \left[\Gamma  -\frac{B^2}{A} \right]}
\sqrt{\frac{\pi}{2A}} \left[ 1 +  {\rm erf} \left(  \frac{B}{\sqrt{2A} }\right) \right],
\end{equation}
where ${\rm erf}(x)=\frac{2}{\sqrt{\pi}}\int_{0}^{x}e^{-y^2}dy$ is the error function. 
Lastly, it is easy to show that the above 
likelihood function corresponds to the following marginalized 
${\tilde \chi}^2_{H}$ function:
\begin{equation}
\label{eq:marginalization}
{\tilde \chi}^{2}_{H}(\phi^{\mu+1})=
\Gamma  -\frac{B^2}{A}
+  \ln A
- 2 \ln \left[ 1 +  {\rm erf} \left(  \frac{B}{\sqrt{2A} }\right) \right]\;.
\end{equation}
where we have ignored the constant $ {\rm ln}(\pi/2)$, since it 
does not play a role 
during the minimization procedure. 

Obviously, the statistical estimator (\ref{eq:marginalization})
does not suffer from the Hubble constant
problem. Indeed, instead of minimizing $\chi^{2}$  
we now use the marginalized 
${\tilde \chi}^2_{H}$ function which
is independent of $H_{0}$
and thus we do not need to impose in the statistical analysis an
a priori value for the Hubble constant, as usually done in many 
other studies of this kind. 

Bellow, we test the performance 
of the current statistical procedure at the expansion level 
using some well known dark energy models.

\section{Fitting models to $H(z)$ data}
\label{Cosmology}
In this section we present the expansion rate of the Universe 
in the context of the most basic 
DE models whose 
free parameters are constrained following the procedure of the 
previous section. 
Due to the fact that the $H(z)$ data are well inside in the 
matter dominated era we can neglect the radiation term 
from the Hubble expansion.

Let us now briefly discuss the cosmological models explored in 
the present study.

\begin{itemize}

\item Non-flat $\Lambda$CDM model. 
In this case the Hubble parameter is given by 
\begin{eqnarray}
E(z,\phi^{\mu+1}) = \left[\Omega_{m0} (1+z)^3+\Omega_{\Lambda 0} + \Omega_{K0}(1+z)^2 \right]^{1/2},
\end{eqnarray}
where $\Omega_{K0}$ is the dimensionless curvature density parameter 
at the present time which is defined 
as $\Omega_{K0} =1 -\Omega_{m0}- \Omega_{\Lambda 0}$, hence the cosmological vector 
takes the form $\phi^{\mu+1}=(\Omega_{m0},\Omega_{\Lambda 0})$.


\item wCDM model. 
In this spatially flat model the equation of state parameter 
$w_{d}=p_{d}/\rho_{d}$ is constant \cite{wCDM}, where $\rho_{d}$ is the density 
and $p_{d}$ is the pressure of the dark energy fluid
respectively. Under the latter conditions 
the normalized Hubble function is 
\begin{eqnarray}
E(z,\phi^{\mu+1}) = 
\left[\Omega_{m0} (1+z)^3+\Omega_{d0}(1+z)^{3(1+w)}\right]^{1/2},
\end{eqnarray}
where $\Omega_{d0}=1-\Omega_{m0}$ and thus the cosmological vector 
is $\phi^{\mu+1}=(\Omega_{m0},w)$.

\item CPL model. This cosmological model was first introduced in the 
literature by Chevalier-Polarski-Linder 
\cite{Linder03}, \cite{ChevallierPolarski}. Here the equation 
of state parameter is allowed to vary with redshift and it 
is written as a first order Taylor expansion
around the present epoch, $w(a)=w_{0}+w_{1}(1-a)$ with $a=1/(1+z)$. 
Therefore, the dimensionless Hubble parameter takes
the following form
\be
 E(z,\phi^{\mu+1}) = \left[\Omega_{m0}(1+z)^3+\Omega_{d 0}X(z) \right]^{1/2},
\ee
\end{itemize}
where
$$
X(z)=(1+z)^{3(1+w_{0}+w_{1})}{\rm exp}\left(-3w_{1}\frac{z}{z+1} \right) 
$$
and $\Omega_{d0}=1-\Omega_{m0}$. In this case the vector of the model
parameters is $\phi^{\mu+1}=(\Omega_{m0},w_{0},w_{1})$.

For the non-flat $\Lambda$CDM model the likelihood function peaks at
$(\Omega_{m0},\Omega_{\Lambda 0})=(0.250^{+0.039}_{-0.043},0.693^{+0.147}_{-0.186})$ with
${\tilde \chi^2}_{H,\rm min}/df\simeq 0.639$ ($df$ are the degrees of
freedom). Also, based on $\Omega_{K0} =1 -\Omega_{m0}- \Omega_{\Lambda 0}$ 
we find $\Omega_{K0} =0.057^{+0.142}_{-0.152}$.
Our constraints are in agreement
within $1\sigma$
errors to those of Farooq et al. \cite{Farooq16} 
who found, using the same $H(z)$ data, 
$(\Omega_{m0},\Omega_{\Lambda 0})= 
(0.23,0.60)$ for $H_{0}=68$ Km/s/Mpc and 
$(\Omega_{m0},\Omega_{\Lambda 0})= 
(0.25,0.78)$ for $H_{0}=73.24$ Km/s/Mpc respectively.
Recently, Jesus et al. \cite{Jesus} found $H_0=69.5 \pm 2.5$Km/s/Mpc, 
$\Omega_{m0}=0.242\pm 0.036$ 
$\Omega_{\Lambda 0}=0.256\pm 0.14$, while
using the Riess et al. \cite{Riess11} prior $H_{0}=73.8$Km/s/Mpc they found 
$0.21\le \Omega_{m0}\le 0.32$ and $0.65\le \Omega_{\Lambda 0}\le 0.99$.

In the case of wCDM cosmological model the results of 
the minimization analysis are 
$(\Omega_{m0},w)=(0.262^{+0.042}_{-0.037},-0.96^{+0.275}_{-0.270})$ with
${\tilde \chi}^2_{\rm min}/df\simeq 0.64$. For comparison 
Farooq \emph{et al.} \cite{Farooq16} obtained 
$(\Omega_{m0},w_{0})= 
(0.26,-0.86)$ for $H_{0}=68$ Km/s/Mpc and 
$(\Omega_{m0},w_{0})= 
(0.24,-1.06)$ for $H_{0}=73.24$ Km/s/Mpc respectively.
Lastly, for the CPL parametrization 
we find: ${\tilde \chi}^{2}_{\rm min}/df \simeq 0.64$ and
$(w_{0},w_{1})=(-0.960\pm{0.171},0.047\pm{0.425})$, where we have set 
$\Omega_{m0}=0.262$.  We repeat our analysis by using the 
$\Omega_{m}$-prior derived originally by the Planck team 
\cite{Planck16}. Specifically, if we impose $\Omega_{m0}=0.308$ then 
we obtain $(w_{0},w_{1})=(-0.687\pm{0.123},-1.009\pm{0.598})$
with ${\tilde \chi}^{2}_{H,\rm min}/df \simeq 0.66$.
Notice, that in Table II we provide a more compact presentation 
of our statistical results.
In Fig.~2 we plot the 1$\sigma$, 2$\sigma$ and $3\sigma$
confidence contours in the $(\Omega_{m0},\Omega_{\Lambda 0})$ 
and $(\Omega_{m0},w)$ planes for non-flat $\Lambda$CDM
(upper panel) and wCDM 
(bottom panel) models respectively.
We observe that our best-fit values are almost $\sim 1\sigma$ 
away, from the values provided by the Planck team \cite{Planck16}
(see stars in Fig. 2). 
Moreover, in Fig.~3 we show the $(w_{0},w_{1})$ contours for the CPL
model by using $\Omega_{m0}=0.262$ (upper panel) and 
$\Omega_{m0}=0.308$ (bottom panel). The stars in
Fig.~3 corresponds to the solution $(w_{0},w_{1})=(-1,0)$.
As expected, we find that the parameter $w_{0}$ is degenerate
with respect to $w_{1}$, implying that 
the time varying equation-of-state parameter $w(z)$ 
is not constrained by this analysis. 

\begin{figure}[ht]
\label{results1}
\includegraphics[width=0.5\textwidth]{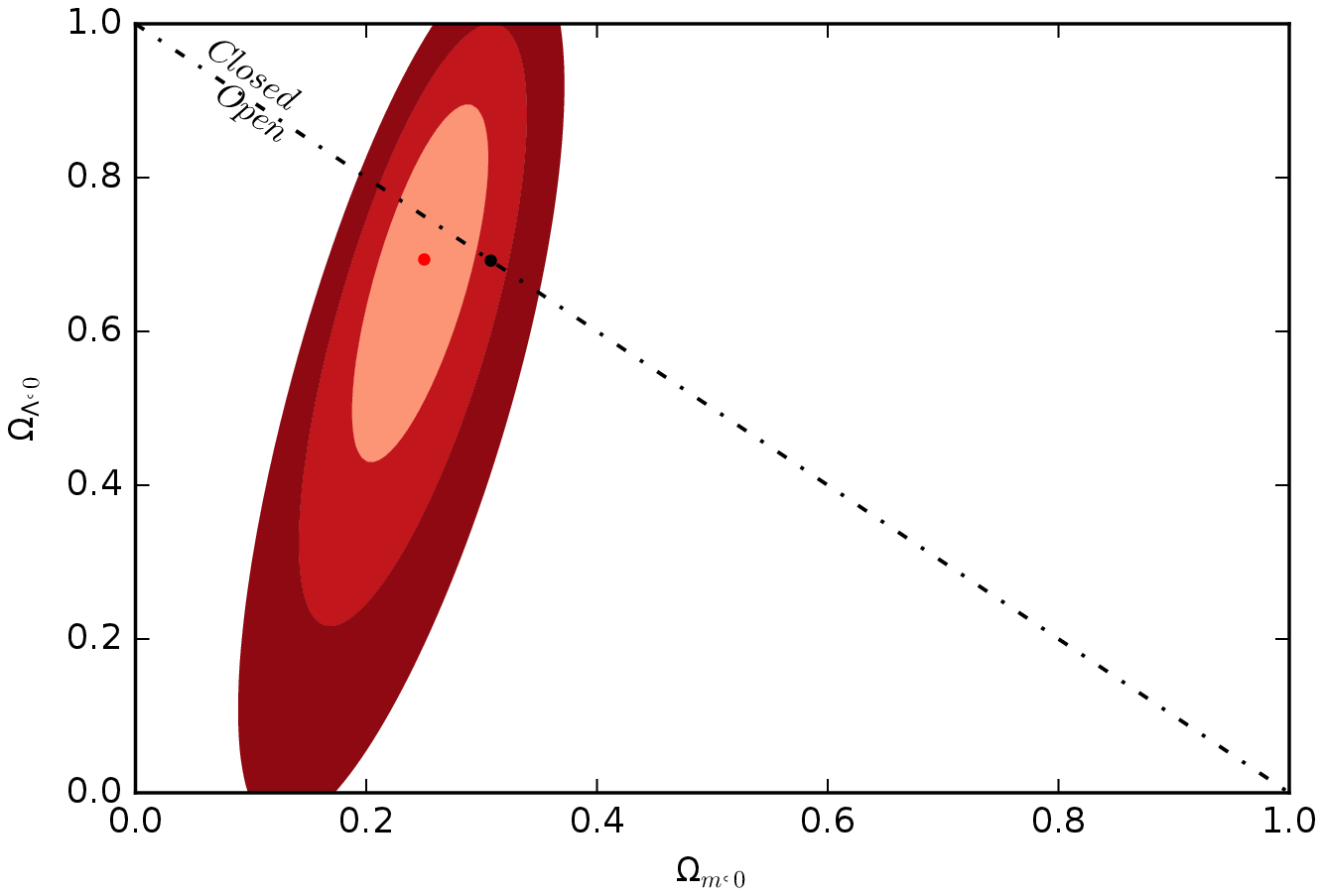}
\includegraphics[width=0.5\textwidth]{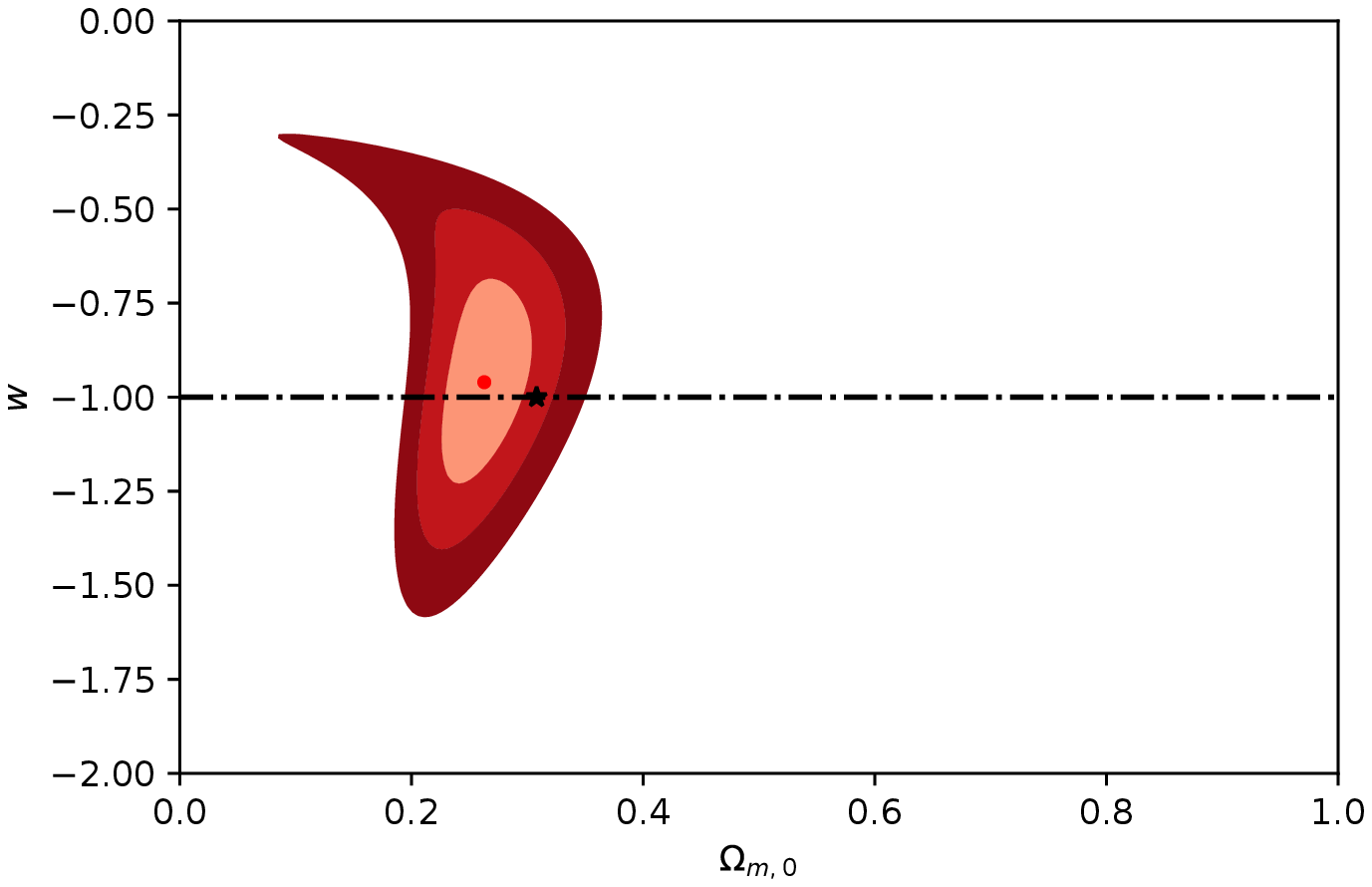}
\caption{The likelihood
contours for $\Delta {\tilde \chi}^2={\tilde \chi}^{2}_{H}-{\tilde \chi}^{2}_{H, \rm min}$ 
equal to 1$\sigma$ (2.32), 2$\sigma$ (6.18) and 3$\sigma$ (11.83) confidence levels. The red dot corresponds to the best-fit solutions. {\it Upper panel:} 
the contours of the non-flat \lcdm model, in the $(\Omega_{m0},\Omega_{\Lambda})$ plane. The dashed line represents the $\Omega_{m0}+\Omega_{\Lambda}=1$ line. 
Here the best fit point is 
$(\Omega_{m0},\Omega_{\Lambda})=(0.250,0.693)$.
{\it Lower panel:} the wCDM model in the $(\Omega_{m0},w)$ plane. 
The best fit solution is $(\Omega_{m0},w)=(0.262,-0.960)$. The dashed curve 
corresponds to $w = -1$. Notice, that
stars show the best-fit 
solution provided by the Planck team, \cite{Planck16} for 
the flat $\Lambda$CDM model.}
\end{figure}

\begin{figure}[ht]
\label{results2}
\includegraphics[width=0.5\textwidth]{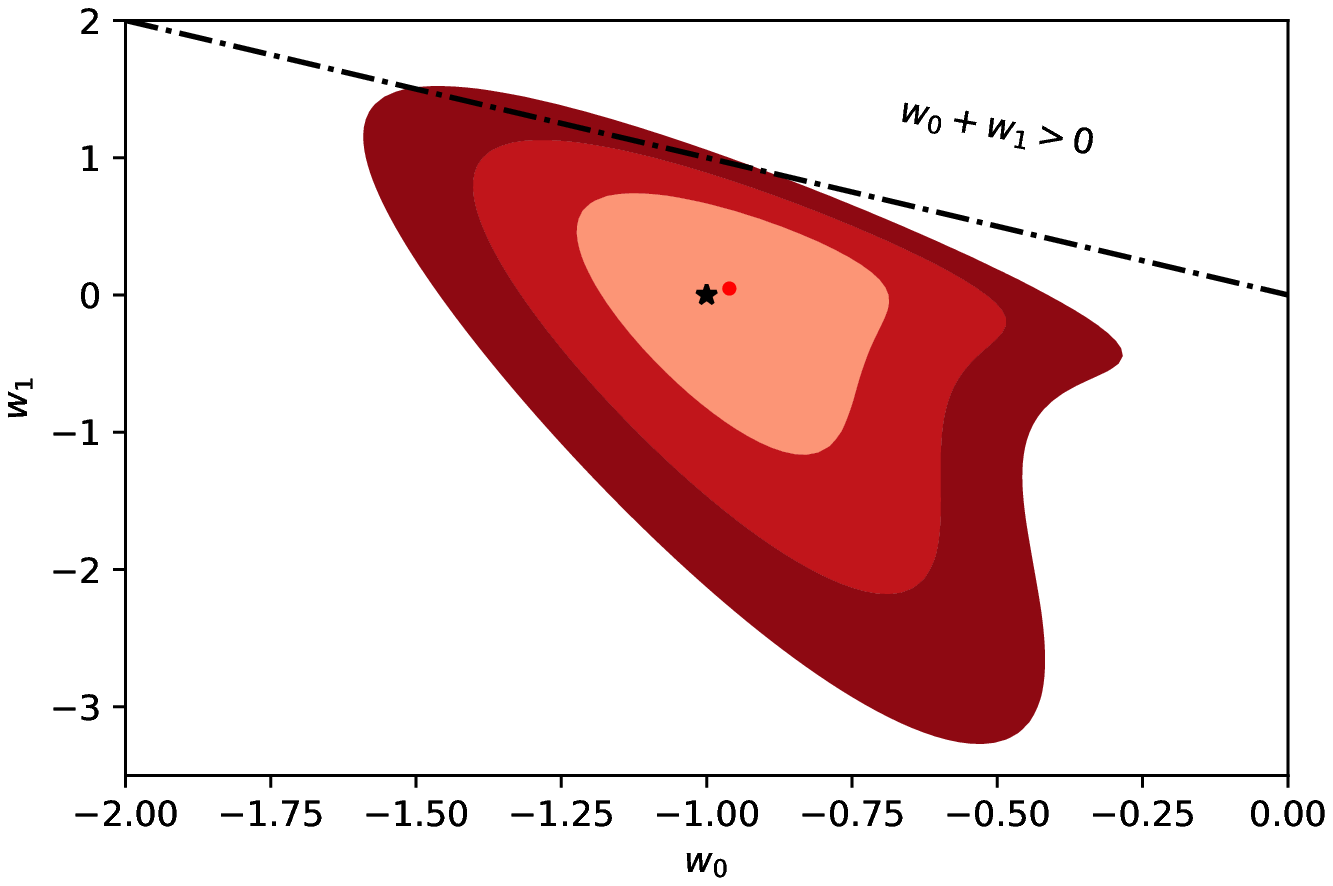}
\includegraphics[width=0.5\textwidth] {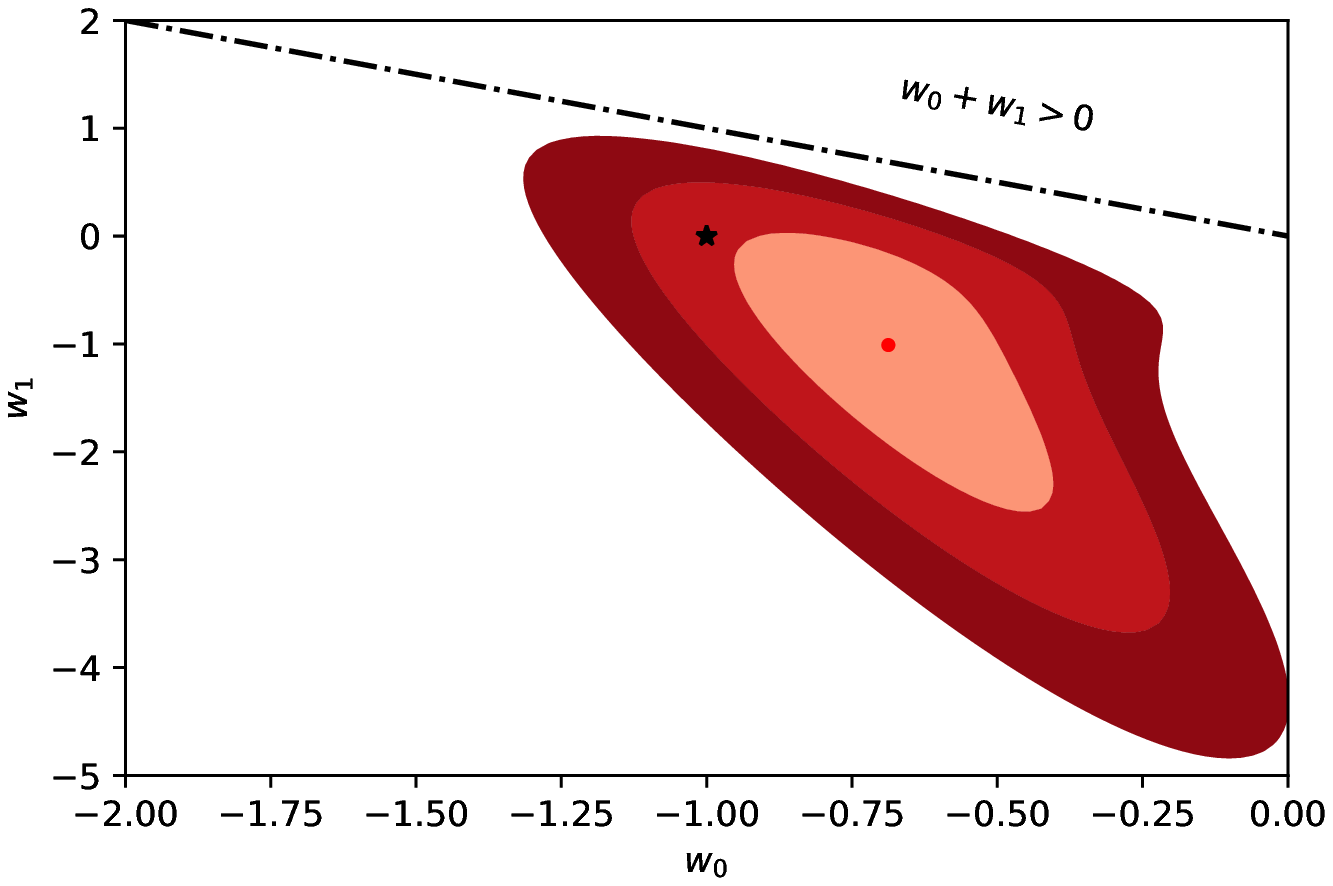}
\caption{The likelihood
contours $\Delta {\tilde \chi}^2={\tilde \chi}^{2}_{H}-{\tilde \chi}^{2}_{H, \rm min}$ in 
the case of CPL model.
{\it Upper panel:} Here we utilize 
$\Omega_{m0}=0.262$ from the first panel of Table II. 
{\it Bottom panel:} Here we use  
$\Omega_{m0}=0.308$ from Planck, \cite{Planck16}.
Notice that stars corresponds to flat $\Lambda$CDM model $(w_{0},w_{1})=(-1,0)$.
}
\end{figure}

\begin{figure}[ht]
\label{results3}
\includegraphics[width=0.5\textwidth]{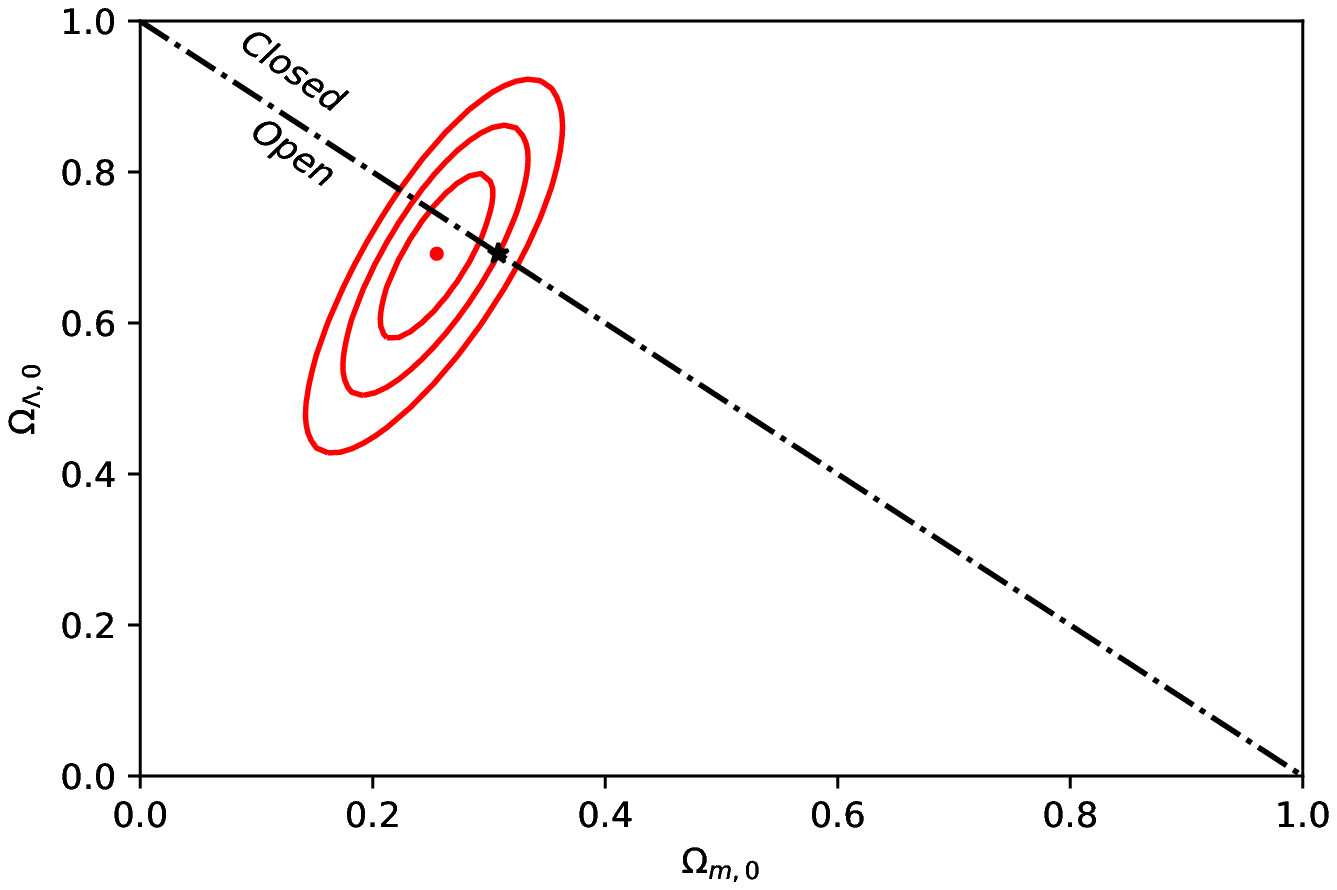}
\includegraphics[width=0.5\textwidth] {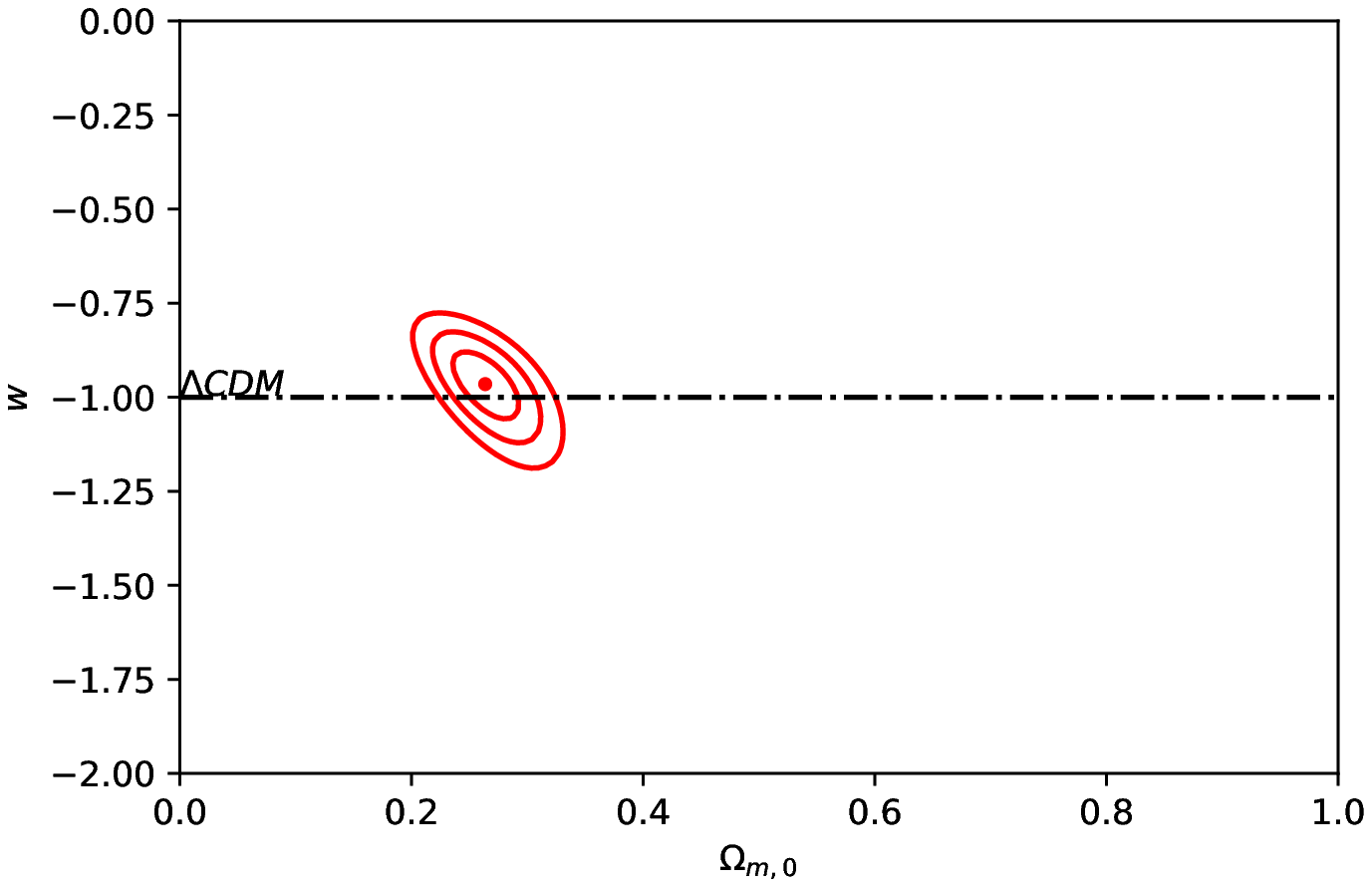}
\caption{The $H(z)$/SNIa joint likelihood
contours. 
The upper panel shows the solution space for the non-flat  
\lcdm model, while the lower panel corresponds to 
wCDM model. The dashed line corresponds to $w = -1$.
The red dot corresponds to the best-fit solutions.
The black star shows the solution of Planck \cite{Planck16}.}
\end{figure}

\begin{figure}[ht]
\label{results4}
\includegraphics[width=0.5\textwidth]{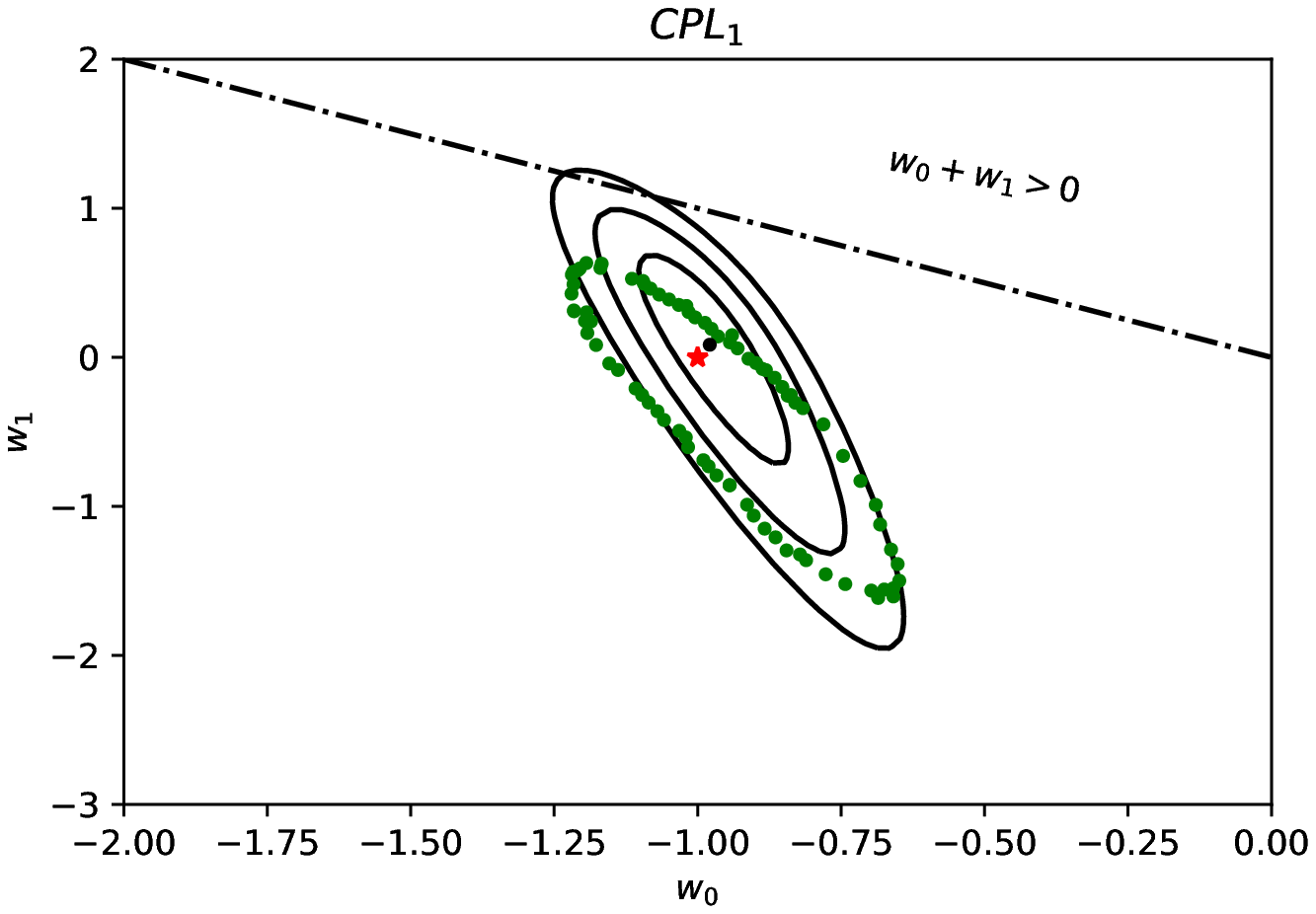}
\includegraphics[width=0.5\textwidth] {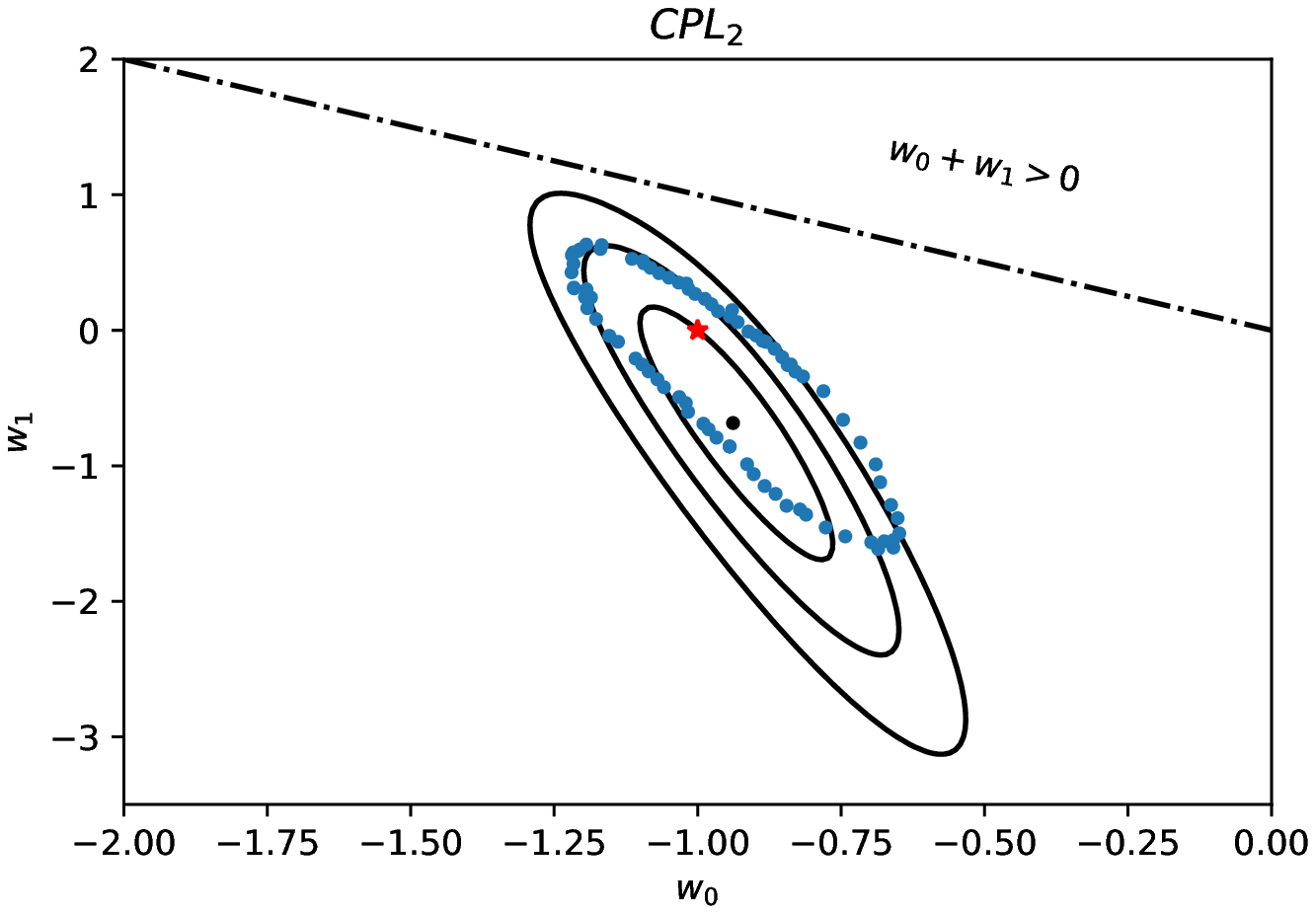}
\caption{The joint SNIa and $H(z)$ 
  likelihood contours 
in the ($w_{0},w_{1}$) plane
for $\Omega_{m0}=0.262$ (upper panel) and 
$\Omega_{m0}=0.308$ (lower panel). 
The solid black dots 
correspond to the best fit parameters. We also
show the theoretical $\Lambda$CDM
$(w_{0},w_{1})=(-1,0)$ values (star points). The dot-dashed line 
corresponds to $w_{0}+w_{1}=0$.
Finally, the area of 
green/blue dots borrowed from
Planck \cite{Planck16}.
}
\end{figure}

\subsection{Joint Analysis with SN Ia}
Although the $H(z)$ data provide a direct measurement
of the expansion of the Universe, due to their large errors with respect 
to the SN Ia data, 
various authors preferred to utilize the latter data 
in order to constrain the cosmological parameters\footnote{For an thorough treatment of the statistical difficulties see Ref. \cite{Barbary}}. 
Here we want to combine $H(z)$ and SN Ia in order to study 
the performance of the $H(z)$ data (as they stand today, namely 38 entries) 
with that of SN Ia data.
In particular, we use the {\em Union 2.1} set of 580 SN Ia of
Suzuki \emph{et al}. \cite{Suzuki:2011hu}.
Concerning the chi-square 
estimator of the SN Ia we utilize the method of \cite{Nes05}, where
the form of ${\tilde \chi}^{2}_{\rm Sn}$ is independent of $H_{0}$ (see 
also Ref.\cite{Chavez2016} and references therein).  
In this framework, the overall likelihood function is given by 
the product of the individual likelihoods according to:
$$
 \cal L_{\rm tot}={\cal L}_{\rm sn} \times {\cal L}_{H} 
$$
which translates in an addition for the total $\chi^2_{\rm tot}$: 
$$
 \chi^2_{\rm tot}={\tilde \chi}^2_{\rm sn}+{\tilde \chi}^2_{H}
$$

The results based on the joint analysis of $H(z)$/SNIa data are given in 
Figs. (4-5) and listed in the second panel of Table II.
It becomes clear that the addition of the SNIa data in the 
likelihood analysis improves substantially the statistical results.
Overall, we find that the $H(z)$/SNIa joint 
analysis increases the Figure of Merit 
(FoM: for definition see below) by a factor of $\sim 2.5$ 
with respect to that of $H(z)$ analysis.
Therefore, 
the combined analysis of the $H(z)$ data with SNIa 
reduces significantly the parameter
space, providing tight constraints on the non-flat $\Lambda$CDM and wCDM models 
respectively. In particular, for the former model the total likelihood 
function peaks at 
$(\Omega_{m0},\Omega_{\Lambda 0})=(0.255\pm{0.02},0.692\pm{0.045})$ with $\chi^2_{\rm tot,min}/df\simeq 0.950$, while 
for the latter cosmological model we find 
$(\Omega_{m0},w)=(0.264\pm{0.015},-0.965\pm{0.046})$ with $\chi^2_{\rm tot,min}/df\simeq 0.950$. 
Concerning the 
CPL model we find that 
although the area of
$w_{0}-w_{1}$ contours is significantly reduced,
the degeneracy between $w_{0}$ and $w_{1}$ persists also in 
the joint analysis. 
However, what is specifically interesting 
is that for the CPL model the $H(z)$/SNIa contours are in very good agreement 
with those of Planck TT, lowP CMB data and external (BAO, JLA, $H_0$) data, \cite{Planck16} (see solid circles in Fig.5), which confirms that our 
analysis with the $H(z)$ and SNIa probes correctly reveals 
the expansion history of the Universe as provided by the Planck team.


Concluding this section it is interesting to mention that recently, 
Yu {\it et al.} \cite{Yu17} 
introduced the covariance matrix of 
three BAO $H(z)$ measurements \cite{Alam2016} in the 
$H(z)$ analysis. Using this covariance matrix we have re-done 
our statistical analysis and in Table III we provide the corresponding 
constraints, which are in agreement (within 1$\sigma$ errors)  
with those of Table II.
Notice, that in the appendix 
we have generalized the statistical methodology of section II 
in the presence of the covariance matrix.

\section{Strategy to improve the cosmological constrains using the $H(z)$ data}
\label{Strategy}
From the previous analysis it becomes clear that an important question
that we need to address is the following: 
{\it What is the 
strategy for the
recovery of the dark energy equation of state using 
the direct measurements of the Hubble expansion? }
In this section we proceed with our investigation towards 
studying the effectiveness of utilizing 
$H(z)$ measurements to constrain the equation of state parameter.
Specifically, our aim is 
to test how better can we go in placing cosmological constraints
by increasing the current $H(z)$ sample from 38 to 100.
In order to achieve such a goal, we produce sets of 
Monte Carlo simulations with which we quantify our ability to 
recover the input cosmological parameters of a fiducial 
cosmological model, namely $(\Omega_{m0},\Omega_{K0},w_{0},w_{1})=(0.25,0,-1,0)$
with $H_{0}=68.75$~Km/s/Mpc. 
In the upper panel of Fig.~6 we present the 
evolution of the Hubble parameter of the reference model (see solid) line
and on top of that we plot the $H(z)$ data (solid points). 
In the lower panel of 
Fig ~6 we show the distribution of 
$100(\%)\times|H_{D}-H_{\rm ref}|/H_{D}$ 
as a function of redshift (see below), where $H_{D}$ and $H_{\rm ref}$ 
are the Hubble parameters of the data and the reference cosmology
respectively. 
We verify that the differences 
$\delta H=|H_{D}-H_{\rm ref}|$ are not correlated with redshift.

\begin{figure}[ht]
\includegraphics[width=0.5\textwidth]{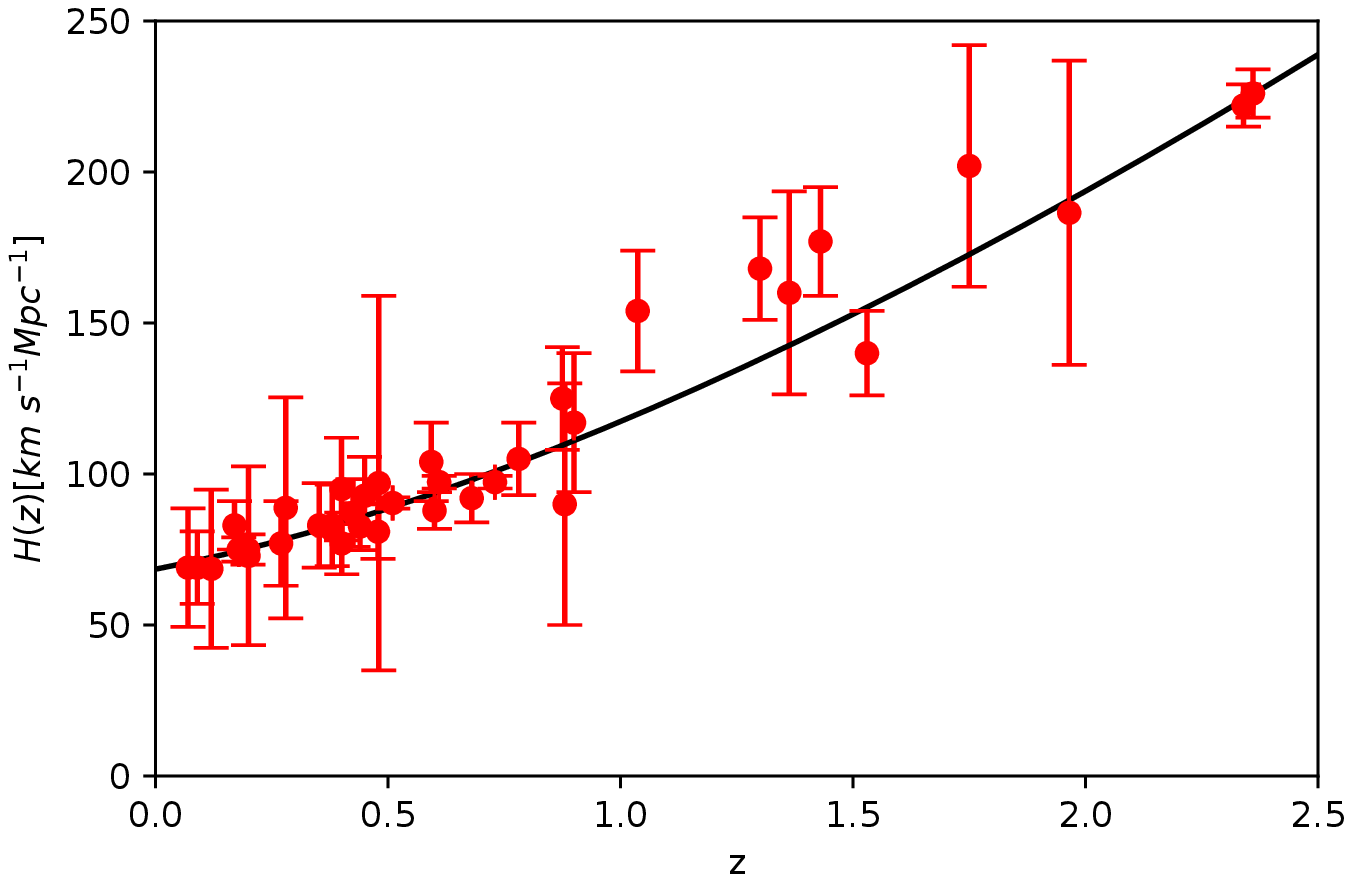}
\includegraphics[width=0.5\textwidth]{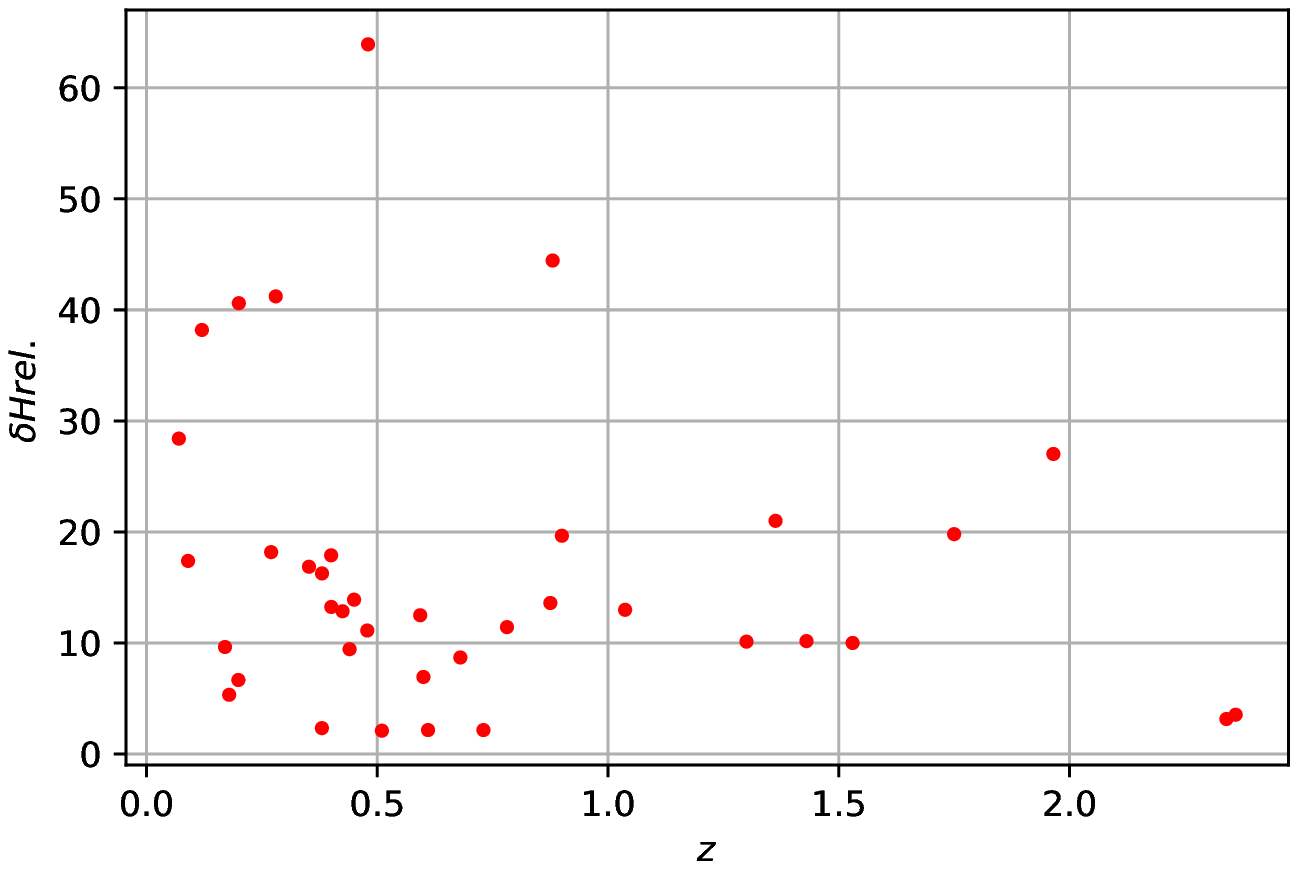}
\caption{{\it Upper panel:}
Comparison of the observed (red points~\cite{Farooq16})
and theoretical evolution of the reference Hubble parameter,
$H(z)$, using $(\Omega_{m0},\Omega_{K0},w_{0},w_{1})=(0.25,0,-1,0)$ and 
$H_{0}=68.75$Km/s/Mpc. The reference cosmology is 
represented by the solid curve. 
{\it Lower panel:} The distribution of $\delta H=|H_{D}-H_{\rm ref}|$. 
Notice, that $H_{D}(z)$ indicates the observed 
Hubble parameter, while $H_{\rm ref}(z)$ is the Hubble 
function of the fiducial cosmology.
}\label{fig:hubble}
\end{figure}

Now, we develop an algorithm that generates different number of 
mock $H_{\rm MC}(z)$ measurements following the redshift and the error 
distributions of the real $H(z)$ data (see Figures 1 and 2).
Thus, our aim is to obtain the value of $H_{\rm MC}$ as well as the
corresponding $1\sigma$ error by calibrating the mock $H(z)$ 
sample from the real $H(z)$ data in which $0.07\leq z\leq
2.36$. More specifically, we implement the following steps:

First, from the redshift interval $[0.07,2.36]$ we choose a redshift
$z_{\rm ran}$ 
by randomly sampling the observed redshift distribution (see Fig.1).
For this ''random'' redshift we define the 
measured Hubble parameter $H_{D}(z_{\rm ran})$ 
and the ideal Hubble parameter $H_{\rm ref}(z_{\rm ran})$ 
from the reference cosmology. Second, in order to take into account
the deviation of the observed Hubble parameter from the reference cosmology
we are randomly sampling 
the distribution of the 
differences $\delta H$ (see lower panel of Fig. 6) 
between the data and the fiducial cosmological model.
Once, steps (1) and (2) are completed for all mock data\footnote{We sample 
the number of mock data as follows $N\in [38,120]$ in steps of 2.} used, the 
mock Hubble parameter $H_{\rm MC}$ is selected 
from the normal distribution 
${\cal N}(H_{\rm ref},\sigma^{2}_{\rm ran})$. Finally, 
performing a trial and error procedure we have confirmed that by 
assigning 
to each mock Hubble parameter $H_{\rm MC}$ the individual error 
$\sigma_{\rm ran}=\sqrt{\sigma_{H}^{2}+\delta H^{2}}$ we recover the 
contours of the reference model and thus the mock $H(z)$ data 
contain the following simulated triads 
$ \{ z_{\rm ran},H_{\rm MC},\sigma_{\rm ran} \} _{i} $, where $i=1,..N$ and 
$N\in [38,120]$. 
For the benefit of the reader in Fig.7 we plot 
the mock Hubble parameter as a function of redshift. Notice that 
in this case the mock sample constraints $N=100$ entries.
This figure can be compared with that of the observed $H(z)$ data 
(see upper panel of Fig. 6).

\begin{figure}[ht]
\includegraphics[width=0.5\textwidth]{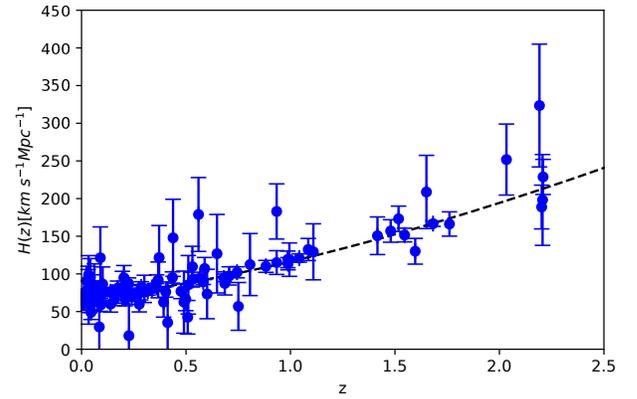}
\caption{
The mock Hubble parameter as a function of redshift. 
In this case the mock data-set constraints $N=100$ entries. The dashed line corresponds to \lcdm model with $(H_{0},\Omega_{m0},\Omega_{\Lambda 0}) = (68.5,   0.25,   0.693) $.
}\label{fig:hubblemock}
\end{figure}

Now, based on the mock data we attempt 
to measure the effectiveness of the $H(z)$ measurements in 
constraining the cosmological 
parameters. Therefore, we calculate the well known 
Figure-of-Merit (FoM) in the 
solution space. The FoM is a useful tool because it 
provides an 
assess how constraining the
likelihood analysis of the $H(z)$ data can be. 
We have defined the FoM
as the inverse of the enclosed area of the 2$\sigma$ 
contour in the parameter space 
of any two degenerate cosmological parameters, namely  
$\Omega_{m 0}-\Omega_{\Lambda 0}$ and $w_{0}-w_{1}$.
Of course, the higher the FoM is, the more constraining the model.
We generate 100 Monte-Carlo simulations for each selected
number ($N=38,40,..120$) of mock $H(z)$ data, and the corresponding 
results are shown in Fig. 8. In this figure we plot the ratio
between the simulation FoM and that of the present sample of 
38 $H(z)$ measurements, namely FoM$_{38}$, as a function of the number of mock 
$H(z)$ data. Therefore, with the aid of Figure 8 we see the behavior 
of the factor by which the
FoM increases with respect to its present value.
We observe that this factor
increases linearly with the number of $H(z)$ mock data. 
A linear regression yields

$$
\left( \frac{FoM}{FoM_{38}} \right)_{\rm non-flat, \Lambda}= (0.0087 \pm 0.0002)N + 0.689 \pm 0.027
$$
$$
\left( \frac{FoM}{FoM_{38}} \right)_{\rm CPL}=(0.0246 \pm 0.0007)N -0.534 \pm 0.33
$$
Using the above expressions we find that for the realistic future observations 
of $\sim 100$ $H(z)$ data the FoM 
is expected to increase by a factor of $\sim 2$ and $\sim 3$  
for the non-flat $\Lambda$CDM and CPL models respectively.

\begin{table*}[ht]
\caption[]{Results of cosmological parameters values and uncertainties. 
Here the $H(z)$ data are not correleted.}
\tabcolsep 2.5pt
\begin{tabular}{cccccc} \hline \hline
Mod. & $\Omega_{m0}$ & $\Omega_{\Lambda 0}(\Omega_{de})$ & $w_{0}$ & $w_{1}$ & $\chi^{2}_{min}/\cal \nu$ \\ \hline \\
\multicolumn{6}{c}{Using only the $H(z)$ data}\\
& &  & & & \\
\lcdm & $0.250_{-0.043}^{+0.039}$ & $0.693_{-0.186}^{+0.147}$& -1 & 0 & 0.639\\
${\rm wCDM}$ & $0.262_{-0.037}^{+0.042}$ & $0.738$ & $-0.960_{-0.270}^{+0.275}$ & 0 & 0.640\\ 
${\rm CPL}$ & 0.262 & 0.738 & $-0.960 \pm{0.171}$ & $0.047\pm{0.425}$ & 0.640\\
${\rm CPL}$ & 0.308 & 0.692 & $-0.687\pm{0.123}$ & $-1.009\pm{0.598}$ & 0.657\\
\hline \\
\multicolumn{6}{c}{Using the joint analysis of $H(z)$/SNIa data}\\
& &  & & & \\
\lcdm & $0.255 \pm{0.020} $ & $0.692 \pm{0.045} $& -1 & 0 & 0.950\\
${\rm wCDM}$ & $0.264 \pm{0.015}$ & $0.736$ & $-0.965\pm{0.046}$ & 0 & 0.950\\ 
${\rm CPL}$ & 0.264 & 0.736 & $-0.979\pm{0.260}$ & $0.085\pm{0.094}$ & 0.950\\
${\rm CPL}$ & 0.308 & 0.692 & $-0.938 \pm{0.053} $ & $-0.684 \pm{0.288} $ & 0.955\\ 

\end{tabular}
\end{table*}

\begin{table*}[ht]
\caption[]{Cosmological constraints using the correlation matrix of the $H(z)$ data
\cite{Alam2016,Yu17}.} 
\tabcolsep 2.5pt
\begin{tabular}{cccccc} \hline \hline
Mod. & $\Omega_{m0}$ & $\Omega_{\Lambda 0}(\Omega_{de})$ & $w_{0}$ & $w_{1}$ & $\chi^{2}_{min}/\cal \nu$ \\ \hline \\
\multicolumn{6}{c}{Using only the $H(z)$ data}\\
& &  & & & \\
\lcdm & $0.255 \pm{0.026} $ & $0.692 \pm{0.142} $& -1 & 0 & 0.747\\
${\rm wCDM}$ & $ 0.248 \pm{0.024}$ & $0.752$ & $-1.015\pm{0.177}$ & 0 & 0.750\\
${\rm CPL}$ & 0.248  & 0.752 & $-1.011\pm{0.332}$ & $-0.110\pm{0.122}$ & 0.749\\
${\rm CPL}$ & 0.308 & 0.692 & $-0.565 \pm{0.221} $ & $-1.564 \pm{0.731} $ & 0.738\\ 
\hline \\
\multicolumn{6}{c}{Using the joint analysis of $H(z)$/SNIa data}\\
& &  & & & \\
Mod. & $\Omega_{m0}$ & $\Omega_{\Lambda 0}(\Omega_{de})$ & $w_{0}$ & $w_{1}$ & $\chi^{2}_{min}/\cal \nu$ \\ \hline
\lcdm & $0.248 \pm{0.016} $ & $0.701 \pm{0.065} $& -1 & 0 & 0.958\\
${\rm wCDM}$ & $ 0.257 \pm{0.005}$ & $0.743$ & $-0.954\pm{0.005}$ & 0 & 0.957\\
${\rm CPL}$ & 0.257  & 0.748 & $-0.946\pm{0.096}$ & $-0.106\pm{0.362}$ & 0.957\\
${\rm CPL}$ & 0.308 & 0.692 & $-0.761 \pm{0.114} $ & $-1.052 \pm{0.551} $ & 0.969\\ 

\end{tabular}
\end{table*}

\begin{figure}[t]
\includegraphics[width=0.5\textwidth]{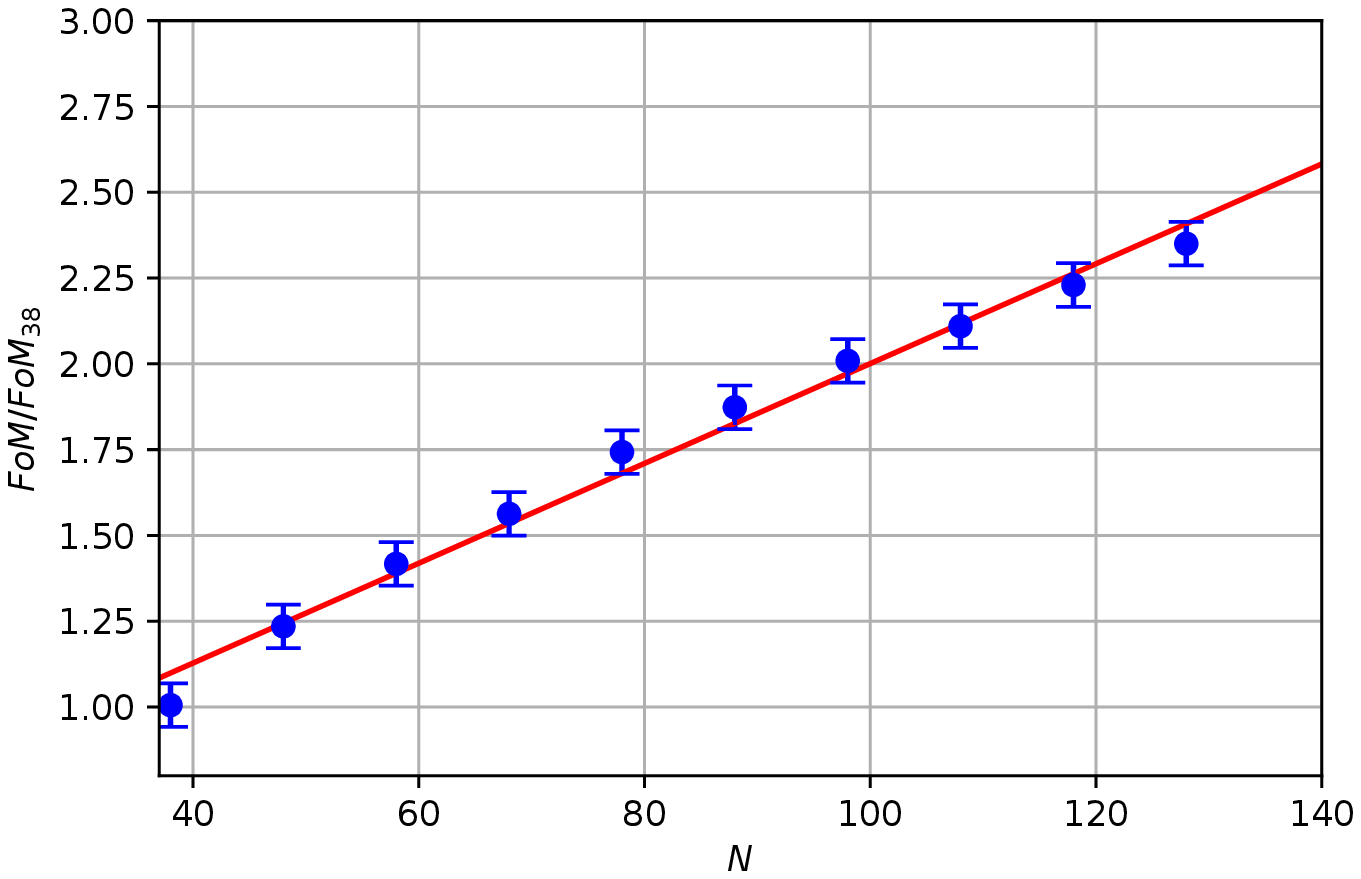}

\includegraphics[width=0.5\textwidth]{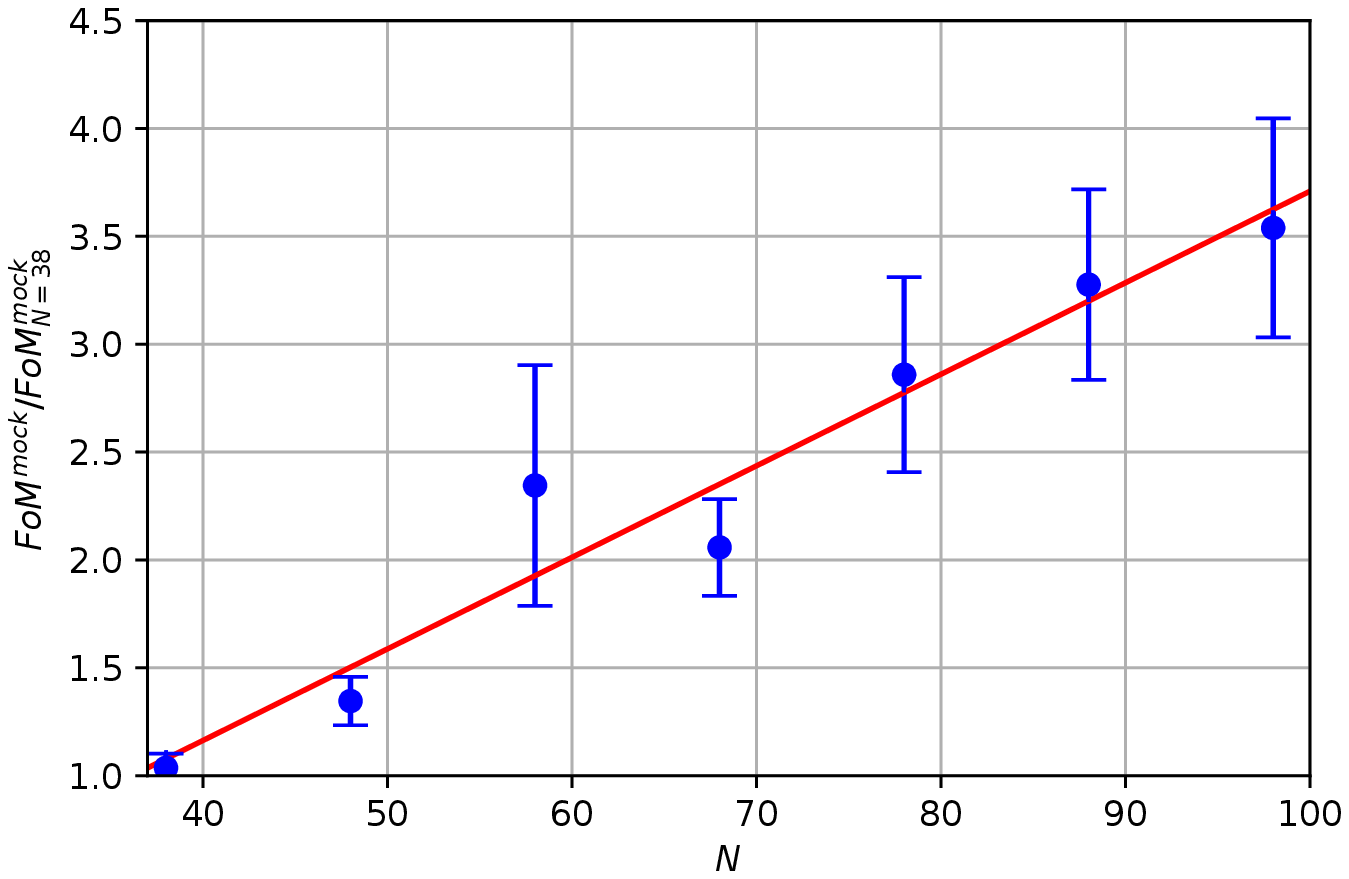}

\caption{The $FoM/FoM_{38}$ as a function of the number of entries 
in the mock $H(z)$ data. Notice, that we used 
100 realizations. The quantity $FoM_{38}$ is the Figure-of-Merrit of the current $H(z)$ data. The upper and lower panels correspond to 
non-flat \lcdm and CPL models respectively.}
\label{results6}
\end{figure}

\section{Discussion}
\label{Discussion}
In this section we provide a qualitative discussion of 
of our $H(z)$ based analysis, giving the reader the 
opportunity to appreciate the new results of our study.
First of all, to our knowledge, this is the first time that 
a proper Bayesian
likelihood analysis applied on the $H(z)$ data 
towards taking out the value of $H_0$ from the 
likelihood analysis. But why is this important 
in this kind of studies? 
Using the direct measurements 
of the cosmic expansion, namely $H(z)$ data 
in constraining the cosmological models, 
via the standard $\chi^{2}$ estimator [see Eq.(\ref{eq:xtetr})], 
one has to either know the exact value of the Hubble constant 
or having it as a free parameter, increasing however the parameter space. 
If we follow the first path then we are facing the well known 
Hubble constant problem. This problem is related 
with the fact that the determination of the
Hubble constant has indicated a $\sim 3.1 \sigma$ tension 
between the value obtained by the Planck team (see \cite{Planck16}), namely   
$H_{0} = 67.8 \pm 0.9$ Km/s/Mpc and the results provided by the  
SNIa project (Riess \emph{et al}. \cite{Riess16}) of 
$H_{0} = 73.24 \pm 1.74$ Km/s/Mpc.
This is the main reason that various studies in the literature 
first imposed the Hubble constant to the above values and then 
they placed constraints
to other cosmological parameters 
($\Omega_{m},\Omega_{\Lambda}, w,..)$. 
For example, Farooq \emph{et al.} \cite{Farooq16} 
provided two different sets 
of constraints for different values of $H_{0}$. Indeed, if they 
imposed $H_{0}=68$ Km/s/Mpc then their likelihood function peaks 
at $(\Omega_{m0},\Omega_{\Lambda 0})= (0.23,0.60)$, while 
for $H_{0}=73.24$ Km/s/Mpc the corresponding likelihood function 
peaks at a different pair, namely 
$(\Omega_{m0},\Omega_{\Lambda 0})= (0.25,0.78)$. 
Obviously, 
the fact that the exact value of the Hubble constant 
remains an open issue in cosmology affects the 
constraints.
At this point we would like to stress that 
our statistical method (see section II) treats in a natural way 
the aforementioned problem. Specifically, the outcome of our 
analysis is a new chi-square estimator [see 
Eq.(\ref{eq:marginalization})] which  
is not affected by the value of the Hubble constant and thus our constraints 
are independent from $H_{0}$.

Concerning the importance of having direct measurements of the cosmic 
expansion some considerations are in order at this point.
The choice of $H(z)$ data, used in many studies the literature as well 
as in our work,
is dictated by the fact that these data are the only data
which are giving a direct measurement of the Hubble expansion 
as a function of redshift.
To date, the cosmic acceleration has been traced mainly  
by SNIa which means that 
the observed 
Hubble relation, namely distance modulus versus $z$, 
lies in the range $0< z< 1.5$ \cite{Suzuki:2011hu,Betoule14}.
In general, the geometrical probes used to map the cosmic
expansion history involve a combination of standard candles 
(SNIa) and standard rulers [clusters,
CMB sound horizon detected through Baryon Acoustic Oscillations
(BAOs; \cite{Blake,Alam2016}) and via the CMB 
angular power spectrum  \cite{Planck16}]. These observations
probe the integral of the Hubble expansion rate $H(z)$, hence they 
give us indirect information of the cosmic expansion 
either up to
redshifts of order $z\simeq 1-1.5$ (SNIa, BAO, clusters) or up to the
redshift of recombination ($z\sim 1100$).
It is therefore clear that the redshift range $\sim 1.5-1000$ is not
directly probed by any of the aforementioned observations, 
and as shown in \cite{Plionis2011} the redshift range $1.5<z<3.5$
plays a vital role in constraining 
the DE equation of state, since different DE models reveal 
their largest differences in this redshift interval. 
Therefore, the fact that direct $H(z)$ measurements 
can be extracted relatively easily at high redshifts
make them, especially those which are  
visible at redshifts $z>1.5$,
indispensable tools towards investigating 
the phenomenon of the accelerated expansion of the universe. 
It is worth mentioning that
there are proposed methods 
which potentially could expand the $H(z)$ measurements to $z\le 5$
\cite{SantageLoeb} (for other possible tracers see \cite{GRB} and 
\cite{Chavez2016}).

At the moment an obvious disadvantage of using alone the 
current $H(z)$ sample in constraining the dark energy models 
is related with the small number statistics and 
thus with the weak statistical constraints. 
However, in order to appreciate the impact of the current 
$H(z)$ data-set in constraining the dark 
energy models, we show in section IIA 
that our combined $H(z)$/SNIa statistical 
analysis (which is not affected by $H_{0})$) 
correctly reveals the expansion of the Universe as provided 
by the team of Planck \cite{Planck16}. 
Specifically, we find that for the CPL model the 
$H(z)$/SNIa contours may compete 
those of Planck TT, lowP CMB data and 
external (BAOs, JLA, $H_0$) data; see solid circles in Fig.5).
In order to understand the effectiveness of $H(z)$/SNIa test
in constraining the $w_{0}-w_{1}$ parameter space, we present 
in the left panel of Fig.9 the $H(z)$ (red-scale contours) 
and {\it Union2.1} SNIa (solid black curves) 
contours respectively. 
We observe that 
even with the current $H(z)$ contours, the joint $H(z)$/SNIa 
analysis reduces 
significantly (due to different inclination of the contours) the $w_{0}-w_{1}$ 
solution space and hence it becomes compatible
to that of Planck (TT, lowP CMB, BAOs, JLA, $H_{0}$) test.

\begin{figure}[ht]
\includegraphics[width=0.5\textwidth]{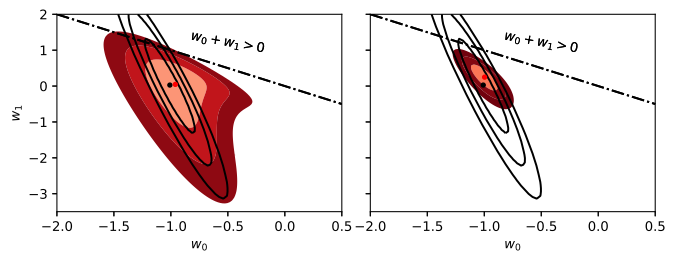}
\caption{{\it Left panel:}
The $1\sigma$, $2\sigma$ and $3\sigma$
likelihood contours in the case of the current $H(z)$ sample, using 
the CPL parametrization.
{\it Right panel:} The corresponding contours in the case of 
our mock sample which contains $\sim 100$ entries.
In black we show the SNIa contours of the 
{\em Union 2.1} set.
}\label{fig:mock}
\end{figure}

Therefore, from the above discussion it becomes clear that 
the ideal avenue that cosmologists 
need to follow towards understanding the nature 
of the cosmic acceleration is to 
use future high quality $H(z)$ data 
to measure the dark energy equation of state 
and the matter content of the Universe. 
This issue is discussed in section IV.
In particular, Monte-Carlo predictions 
show that for the realistic future 
expectations 
of $\sim 100$ $H(z)$ measurements, 
we predict that the present FoM of 
the non-flat $\Lambda$CDM model is increased by a factor of two, while 
in the case of the CPL parametrization 
we find three-fold increase of the corresponding FoM. 
As an example, we provide in the right panel of Fig. 9
the contours of one simulation of 100 $H(z)$ measurements 
for the CPL model in the $w_{0}-w_{1}$ plane (red-scale contours) 
For comparison we plot the corresponding contours (black curves) of the 
{\em Union 2.1} set of 580 SN Ia of
Suzuki \emph{et al}. \cite{Suzuki:2011hu}.
Obviously, 
in the case of SNIa data we observe that the parameters 
$w_{0}$ and $w_{1}$ are degenerate. 
This seems to hold also 
for the JLA data \cite{Betoule14}.
However, our Monte Carlo analysis suggests
that with the aid of only $\sim 100$ future $H(z)$ measurements 
in the redshift range $0<z<2.4$,
we will be able 
to put strong constraints on $w_{0}$ as well as to reduce 
significantly the $w_{1}$ uncertainty and thus testing 
the evolution of the DE equation of state parameter.
We argue that 
having the future $H(z)$ data available we will 
be in a position to use these data combined 
with SNIa and other probes to whittle away the available parameter space 
for the contender dark matter/energy scenarios and hopefully 
to settle on a single viable model.

In a nutshell, we would like to make clear that with the present
analysis we don't want to compete SNIa or other cosmological probes.
The aim of our article is to investigate the power 
of direct measurements of the cosmic expansion, towards 
constraining the dark energy models and to 
provide the appropriate observational framework for future work.

\section{Conclusions}
\label{Conclusions}
We investigated the performance of the latest expansion 
data, the so called $H(z)$ measurements,  
towards constraining the dark energy models. 
In the context of $H(z)$ data aimed
at testing the various forms of dark energy, it
is important to minimize the amount of priors
needed to successfully complete such a task. 
One such prior is the Hubble constant 
and its measurement at the $\sim 1\%$
accuracy level has been proposed as a necessary step
for constraining the dark energy models. 
However, it is well known that the best choice of the value of 
$H_{0}$ is rather uncertain, namely it has 
been found a $\sim 3.1 \sigma$ tension 
between the value provided by the Planck team (see \cite{Planck16})
and the results obtained by the  
SNIa project (Riess \emph{et al}. \cite{Riess16}).
In order to circumvent this problem we implemented in the first 
part of our work a statistical method which is not affected 
by the value of $H_{0}$.  
Based on the latter approach we found that the $H(z)$ data 
do not rule out the possibility 
of either non-flat models or dynamical dark energy cosmological models.

Then we performed a joint likelihood analysis using the $H(z)$ and the 
SNIa data, thereby putting tight constraints on the
cosmological parameters, namely $\Omega_{m0}-\Omega_{\Lambda 0}$ 
(non-flat $\Lambda$CDM model) and 
$\Omega_{m0}-w$ (wCDM model). Furthermore,
using the CPL parametrization 
we found that the $w_{0}-w_{1}$ parameter space provided by 
the $H(z)$/SNIa joint analysis is in a very good agreement 
with that of Planck 2015, which confirms that the present  
analysis with the $H(z)$ and SNIa probes correctly captures 
the expansion of the Universe as found by the team of Planck.

Finally, we performed sets of Monte Carlo simulations in order 
to quantify the ability of the $H(z)$ data to provide strong constraints 
on the model parameters. 
The Monte Carlo approach showed 
substantial improvement of 
the constraints, when increasing the sample to $\sim 100$ $H(z)$ measurements. 
Such a target 
can be achieved in the future, especially in the 
light of the next generation of surveys.

\section*{Appendix}
With the aid of the our statistical method (see section II) 
we calculate the new chi-square estimator that is relevant 
in the case of the covariance matrix.
If the data are correlated then the chi-square estimator is written as:
\begin{equation}
\label{eq:chisq_corr_def}
    \chi^2_{H} = {\bf V} {\bf C}^{-1}_{\rm cov} {\bf V}^{T}, 
\end{equation}
where ${\bf C}^{-1}_{\rm cov}$ is the inverse of the covariance matrix 
\cite{Yu17} and
$$
    {\bf V} = \{H_{\rm obs}(z_1) - H_{M}(z_1,\phi^{\mu}),...,H_{\rm obs}(z_{N}) - 
H_{M}(z_N,\phi^{\mu})\}
$$
or using Eq.(\ref{eq1}) we have 
$$
    {\bf V} = \{H_{\rm obs}(z_1) - H_{0}E(z_1,\phi^{\mu+1}),.., H_{\rm obs }(z_{N} - H_{0}E(z_N,\phi^{\mu+1})\}
$$
Inserting the latter vector into Eq. (\ref{eq:chisq_corr_def}) we obtain 
after some algebra 
$$
    \chi^2_{H} = A H_{0}^2 - 2B H_{0} + \Gamma
$$
and thus following the procedure of section II the functional 
form of the marginalized 
${\tilde \chi}^{2}_{H}$ estimator boils down to that of 
Eq.(\ref{eq:marginalization}). Notice, that the  
quantities $A,B,\Gamma$ are given by
$$
A = {\bf E} {\bf C}^{-1}_{\rm cov} {\bf E}^{T},
$$
$$
B = \frac{1}{2} \left({\bf E} {\bf C}^{-1}_{\rm cov}{\bf H}_{\rm obs}^T+
{\bf H}_{\rm obs} {\bf C}^{-1}_{\rm cov}{\bf E}^T\right)\\
$$
$$
\Gamma = {\bf H}_{\rm obs}{\bf C}^{-1}_{\rm obs}{\bf H}_{\rm obs}^T
$$
with 
$$
{\bf E}=\{E(z_1,\phi^{\mu+1}),...,E(z_N,\phi^{\mu+1}) \}
$$
and 
$$
    {\bf H}_{\rm obs} = \{H_{\rm obs}(z_1),.., H_{\rm obs}(z_{N})\} \;.
$$

\section*{Acknowledgements}
S. Basilakos acknowledges support by the Research Center 
for Astronomy of the Academy of Athens in the
context of the program ''Testing general relativity on cosmological scales''
(ref. number 200/872).


\end{document}